\documentclass[conference,letterpaper]{IEEEtran}
\usepackage[utf8]{inputenc} 
\usepackage[T1]{fontenc}
\usepackage{url}
\usepackage{ifthen}
\usepackage{cite}
\usepackage[cmex10]{amsmath} 
\usepackage{graphicx}

\usepackage{amsfonts} 
\usepackage{amssymb} 
\usepackage{mathtools}

\usepackage{pgfplots}
\usepackage{tikz}
\pgfplotsset{compat=1.17}

\usepackage[algo2e,ruled,vlined,linesnumbered]{algorithm2e}

\usepackage{tikz,pgfplots,pgf}
\usepackage{circuitikz}
\usetikzlibrary{arrows,decorations.pathreplacing}

\usepackage{algorithm}
\usepackage{algpseudocode}

\renewcommand{\leq}{\leqslant}

\renewcommand{\geq}{\geqslant}

\newcommand{\cC}{{\cal C}}

\newcommand{\cH}{{\cal H}}

\newcommand{\cN}{{\cal N}}

\DeclareMathAlphabet{\mathbfsl}{OT1}{ppl}{b}{it} 

\IEEEoverridecommandlockouts 

\pgfplotsset{compat=1.18}
\begin{document}
\title{Erasure Decoding for Quantum LDPC Codes via Belief Propagation with Guided Decimation} 

\author{%
\IEEEauthorblockN{%
    Mert~G\"okduman%
        \IEEEauthorrefmark{1}%
        \IEEEauthorrefmark{2},
    Hanwen~Yao%
        \IEEEauthorrefmark{1}%
        \IEEEauthorrefmark{2},
    and Henry~D.~Pfister%
        \IEEEauthorrefmark{1}%
        \IEEEauthorrefmark{2}%
        \IEEEauthorrefmark{3}%
}
\IEEEauthorblockA{\IEEEauthorrefmark{1}%
                  Duke Quantum Center, Duke University, 
                  Durham, NC, USA
                  }
\IEEEauthorblockA{\IEEEauthorrefmark{2}%
                  Department of Electrical and Computer Engineering, 
                  Duke University, 
                  Durham, NC, USA}
\IEEEauthorblockA{\IEEEauthorrefmark{3}%
                  Department of Mathematics, Duke University, 
                  Durham, NC, USA}
}

\maketitle

\begin{abstract}
Quantum low-density parity-check (LDPC) codes are a promising family of quantum error-correcting codes for fault tolerant quantum computing with low overhead. Decoding quantum LDPC codes on quantum erasure channels has received more attention recently due to advances in erasure conversion for various types of qubits including neutral atoms, trapped ions, and superconducting qubits. Belief propagation with guided decimation (BPGD) decoding of quantum LDPC codes has demonstrated good performance in bit-flip and depolarizing noise. In this work, we apply BPGD decoding to quantum erasure channels. Using a natural modification, we show that BPGD offers competitive performance on quantum erasure channels for multiple families of quantum LDPC codes. Furthermore, we show that the performance of BPGD decoding on erasure channels can sometimes be improved significantly by either adding damping or adjusting the initial channel log-likelihood ratio for bits that are not erased. More generally, our results demonstrate BPGD is an effective general-purpose solution for erasure decoding across the quantum LDPC landscape.
\end{abstract}

\section{Introduction}
\label{sec:intro}

For the development of scalable and fault-tolerant quantum computation, 
quantum error correction is a crucial component that protects quantum information against noise. 
Among the proposed error correction schemes, 
quantum low-density parity-check (LDPC) codes, 
stand out as strong candidates because they promise lower overhead \cite{gottesman2014faulttolerant,fawzi2020constant} 
when compared to topological codes 
such as surface codes \cite{KITAEV20032,dennis2002topological} 
and color codes \cite{bombin2006topological}. 
Belief-propagation (BP) decoding of quantum LDPC codes was first introduced in~\cite{mackay2004sparse} and considered further in~\cite{poulin2008iterative}.
Recent breakthrough results have introduced 
constructions of asymptotically \emph{good} quantum LDPC codes 
with constant rate and linear minimum distance
\cite{panteleev2022asymptotically,leverrier2022quantum,dinur2023good}.
In terms of practical implementation, results by Bravyi et al. 
\cite{bravyi2024high} have shown how certain quantum LDPC codes 
can be embedded into a bilayer hardware architecture.

In this paper, we focus on decoding quantum LDPC codes over 
the quantum erasure channel \cite{bennett1997capacities}.
This model has received more attention recently due to 
proposals and demonstrations 
that erased qubits can be realized 
in several architectures including
neutral atom \cite{wu2022erasure,sahay2023high,ma2023high}, 
trapped ions \cite{kang2023quantum}, 
and superconducting qubits \cite{kubica2023erasure,teoh2023dual}.
Moreover, it has been shown that quantum error correction schemes 
based on erased qubits achieve better finite-length performance,
and higher thresholds compared to those designed for Pauli noise \cite{grassl1997codes,stace2009thresholds,delfosse2021almost,sahay2023high}.
In the erasure model we use, a random subset of the coded qubits is chosen and then subjected to uniform random Pauli errors. While the subset is known to the decoder, the error values are not.

Several decoding algorithms have been proposed for quantum erasure 
correction, targeting various classes of quantum codes in the code. 
In \cite{delfosse2020linear}, a linear-time decoder was introduced 
for surface codes, achieving maximum-likelihood (ML) performance 
by peeling on a spanning tree of erasures on the surface code lattice. 
This was later extended into the union-find decoder, 
initially for topological codes \cite{delfosse2021almost} and subsequently 
for more general quantum LDPC codes \cite{delfosse2022toward}, 
which is capable of correcting both Pauli errors and erasures
with a higher complexity. 
In \cite{lee2020trimming}, a trimming decoder was proposed for 
the erasure decoding of color codes, 
that combines peeling on a spanning tree with erasure set extension 
or vertex inactivation.
Erasure decoding of subsystem color codes has also been studied using 
a combination of techniques, including peeling, clustering, 
and gauge fixing \cite{solanki2021correcting,solanki2023decoding}.

Two erasure decoding algorithms were proposed by Connolly et al.\ in \cite{connolly2024fast}: pruned peeling and vertical-horizontal (VH) decoding.
The pruned peeling decoder can be applied to any code and it combines peeling with a greedy search for stabilizers contained wholly within the erased qubits.
The VH decoder, which can only be applied to hypergraph product (HGP) codes, integrates pruned peeling with an iterative procedure that mitigates vertical and horizontal stopping sets.
For the sub-threshold regime, simulation of the VH decoder has demonstrated performance close to ML with a computational complexity 
of $O(n^2)$ for codes of length $n$.

In this work, we evaluate belief propagation with guided decimation (BPGD) decoding of quantum LDPC codes for the quantum erasure channel.
A recent study by Yao et al.\ demonstrated that
the BPGD decoder has excellent performance for quantum LDPC codes under bit-flip noise and depolarizing noise in the code capacity model~\cite{yao2023belief}. 
Over circuit-level noise, a recent study also shows that a variant of BPGD applied as the inner decoder of sliding window decoding achieves performance on par with BP+OSD for quantum LDPC codes \cite{gong2024lowlatencyiterativedecodingqldpc}.
Here, we show that BPGD outperforms both the peeling and pruned peeling decoders when applied to erasures. While its performance lags slightly behind the VH decoder for HGP codes, BPGD offers a significant computational advantage with its reduced complexity. We further show that tuning the prior log-likelihood ratios (LLRs) and introducing damping techniques significantly improves performance and enables BPGD to approach the performance of VH decoding with reduced complexity.

\section{Preliminaries}
\label{sec:prelim}

\subsection{Classical Erasure Correction}

Error-correcting codes protect information against noise by introducing redundancy.
Let \(\mathbb{F}_2 = \{0, 1\}\) denote the Galois field with 2 elements. A binary linear code \(\mathcal{C} \subseteq \mathbb{F}_2^n\) is a subspace of binary vectors satisfying \(x_1 + x_2 \in \mathcal{C}\) for all \(x_1, x_2 \in \mathcal{C}\). The code can be specified either by a generator matrix \(G \in \mathbb{F}_2^{k \times n}\) whose rows span the code or by a parity-check matrix \(H \in \mathbb{F}_2^{m \times n}\) whose rows are orthogonal to all codewords.

When a codeword is transmitted over the classical binary erasure channel (BEC), each bit is either received correctly or erased with some probability. The received vector \(y \in \{0, 1, ?\}^n\) identifies the positions of the erasures. The decoder's task is to recover the original codeword \(x\)
from the observation $y$. 
In \textit{erasure syndrome decoding}, the decoder only sees the locations of the erasures and the syndrome \(s = H\tilde{y}\), where $\tilde {y} \in \mathbb{F}_2^n$ is a binary version of $y$ whose erasures are replaced by uniform random bits.
This allows the decoder to recover \(x\) if a unique solution exists. For further details, see~\cite[Section 2]{connolly2024fast}.

\subsection{Stabilizer Formalism}

For a single qubit, a pure state is defined by a unit vector in $\mathbb{C}^2$ and the Pauli matrices $I,X,Y,Z$ generate a unitary subgroup that acts on the qubit, where
\[
I = \begin{bmatrix} 1 & 0 \\ 0 & 1 \end{bmatrix}\! , \; 
X = \begin{bmatrix} 0 & 1 \\ 1 & 0 \end{bmatrix}\! , \; 
Y = \begin{bmatrix} 0 & -i \\ i & 0 \end{bmatrix}\! , \; 
Z = \begin{bmatrix} 1 & 0 \\ 0 & -1 \end{bmatrix} \!.
\]
 For an \( n \)-qubit system, the Hilbert space $\cH = (\mathbb{C}^{2})^{\otimes n}$ is used to represent all pure quantum states.
 The Pauli group \( \mathcal{P}_n \) is the subgroup of unitary transformations on $\cH$ that consists of all \( n \)-fold tensor products of the Pauli matrices along with a coefficient in $\{\pm 1,\pm i\}$. 
An element in \( \mathcal{P}_n \) can be written as
\begin{equation}
    P = \alpha P_1 \otimes P_2 \otimes \cdots \otimes P_n,
    \label{}
\end{equation}
\noindent where \( \alpha \in \{\pm 1, \pm i\} \) and \( P_1,P_2,\ldots,P_n \in\{I, X, Y, Z\}\).

 
An $[[n,k]]$ stabilizer code is a quantum error-correcting code that protects $k$ logical qubits encoded into $n$ physical qubits.
A stabilizer group \( \mathcal{S} \) is an abelian subgroup of the Pauli group \( \mathcal{P}_n \) that does not include \(-I^{\otimes n}\).
 The stabilizer code $\cC$ defined by the stabilizer group \( \mathcal{S} \) is  the subspace $\cC \subseteq \cH$ that is invariant under the action of all operators in \( \mathcal{S} \).
 For a stabilizer code, the \textit{stabilizer generators} are a set of operators that generate the stabilizer group \( \mathcal{S} \).

Thus, for any state \( |\psi\rangle \in \mathcal{C} \) and \( S \in \mathcal{S} \), we have:
\begin{equation}
    S|\psi\rangle = |\psi\rangle.
    \label{}
\end{equation}
The \textit{weight} of a Pauli operator \( P \in \mathcal{P}_n \) is the number of qubits on which the Pauli operator acts non-trivially. For instance, the weight of \( X \otimes I \otimes Z \) in $\mathcal{P}_3$ is 2. 
The \textit{distance} \( d \) of a stabilizer code is defined as the minimum weight of a Pauli operator that is not in \( \mathcal{S} \) but which commutes with all elements of the stabilizer group \( \mathcal{S} \). Such an operator is referred to as a \textit{logical operator}, as it acts non-trivially on the encoded qubits.

In this paper, we focus on the \textit{Calderbank-Shor-Steane (CSS) code}, a class of stabilizer code constructed using two classical linear codes \( C_1 \) and \( C_2 \) satisfying \( C_2^\perp \subseteq C_1 \). The stabilizer generators of a CSS code can be divided into \( X \)-type and \( Z \)-type operators:
\[
\mathcal{S}_X = \{X^{\mathbf{v}} : \mathbf{v} \in C_2\}, \quad \mathcal{S}_Z = \{Z^{\mathbf{u}} : \mathbf{u} \in C_1^\perp\},
\]
where $X^{\mathbf{v}} \coloneqq X_1^{v_1} \otimes \cdots \otimes X_n^{v_n}$ and $Z^{\mathbf{u}}$ is defined similarly.
The binary representations of these stabilizers are generated by a pair of parity check matrices, \(H_X\) and \(H_Z\), satisfying
\begin{equation}
    H_X \, H_Z^T = 0.
\end{equation}
This condition is the specialization to CSS codes of the more general \textit{commutativity constraint} required by all stabilizer codes (i.e., that the stabilizer group is commutative).
For CSS codes, the rows of \(H_X\) and \(H_Z\) correspond to \(X\mkern-2mu\)-type and \(Z\)-type  stabilizer generators.
Since stabilizers of the same type automatically commute, the above condition ensures that the stabilizer generators corresponding to \(H_X\) commute with those defined by \(H_Z\).

For noise model where the \(X\mkern-2mu\)- and \( Z\)-type errors are independent, 
this structure simplifies the error correction procedure by leveraging the separation of \( X\mkern-2mu\)- and \( Z\)-type errors. In this context, we cover the correction of \( X\mkern-2mu\)-type errors using classical decoding of \( C_2 \), with the understanding that the correction of \( Z \)-type errors follows the same approach using classical decoding of \( C_1^\perp \), thus effectively reducing the quantum error correction problem to two separate classical decoding tasks.

\subsection{Syndrome Decoding for the Quantum Erasure Channel}

Erasure channels provide a simplified model for information loss by specifying the locations of likely errors. This allows for efficient decoding, especially with LDPC codes, which perform exceptionally well on these channels. In classical coding theory, LDPC codes first demonstrated their potential to approach channel capacity through iterative decoding algorithms like belief propagation \cite{Urbanke2002block, Spielman2001BEC, Pfister2004BEC}. As the study of erasure channels helped advance classical coding, we expect similar benefits for quantum codes in terms of improving decoding strategies and understanding performance limits.

In this work, we consider the quantum erasure channel \cite{bennett1997capacities,grassl1997codes} as our noise model, where each qubit in the encoded state of a stabilizer code is independently erased with probability $p$. 
When a qubit is erased, it is affected by a Pauli operator in $\{I,X,Y,Z\}$ chosen uniformly at random.
The locations of the erasures are given to the decoder as side information.

To correct the Pauli error $E\in\mathcal{P}_n$ affecting the erased qubits, we measure the syndrome of the stabilizers. 
The goal of the decoder is to identify a Pauli error estimate $\hat{E}$ that matches the syndrome and acts trivially outside of the erased locations. 
This decoding process is successful if the decoded error either exactly matches the actual Pauli error or differs from it by a stabilizer. In other words, the decoding is successful if:
\begin{equation}
    E \mathcal{S} = \hat{E} \mathcal{S},
    \label{}
\end{equation}
\noindent where \( \mathcal{S} \) denotes the stabilizer group, and \(E\mathcal{S}\) denotes its coset shifted by \(E\). This condition implies that the identified error is logically equivalent to the actual error up to a stabilizer transformation.

The non-trivial elements of \( E \) are restricted to the locations of the erased qubits. Since the locations of the erasures are known, the decoding problem is constrained to finding an error in the support of the known erasures that matches the syndrome. There are four possible outcomes for this decoding process:

\begin{enumerate}
    \item \textbf{Exact Match:} The decoder identifies an error \( \hat{E} \) that exactly matches the actual error \( E \), meaning \( \hat{E} = E \).
    \item \textbf{Degenerate Match:} The identified error \( \hat{E} \) differs from the error \( E \) by a stabilizer, i.e., \( \hat{E} \cdot \mathcal{S} = E \cdot \mathcal{S} \). In this case, the identified error is logically equivalent to the actual error but does not affect the logical state.
    \item \textbf{Logical Error:} The identified error \( \hat{E} \) differs from the actual error \( E \) by a logical operator, i.e., \( \hat{E} \cdot \mathcal{S} \neq E \cdot \mathcal{S} \), resulting in a logical error that changes the logical state.
    \item \textbf{Decoder Failure:} The decoder fails to find any error in the support of the erasures that matches the syndrome, resulting in no valid decoding solution.
\end{enumerate}

The probability of each Pauli operator \( I \), \( X \), \( Y \), or \( Z \) on the erased qubits is \( 1/4 \), as previously stated. Importantly, for CSS codes, these probabilities translate to independent \( 1/2 \) probabilities for \( X \mkern-2mu \)-type and \( Z \)-type errors. This is because a Pauli-\( Y \) error corresponds to both an \( X \mkern-2mu\)-type error and a \( Z \)-type error occurring on the same qubit.

Thus, the syndrome decoding process is separated into independent corrections of \( X \)-type and \( Z \)-type errors. The syndrome vector \( s_X \) corresponding to \( Z \)-stabilizers provides information about \( X \mkern-2mu \)-errors, and the syndrome vector \( s_Z \) from \( X \mkern-2mu \)-stabilizers provides information about \( Z \)-errors. The decoding problem reduces to solving:
\begin{equation}
    s_X = H_Z \, e_X^T, \quad s_Z = H_X \, e_Z^T,
    \label{}
\end{equation}
\noindent where \( H_X \) and \( H_Z \) are the parity-check matrices corresponding to the \( X \mkern-2mu \)- and \( Z \)-stabilizers, and \( e_X \) and \( e_Z \) are the error vectors for \( X \mkern-2mu \)- and \( Z \)-type errors, respectively.

\subsection{Quantum LDPC Codes}

Quantum Low-Density Parity-Check (QLDPC) codes are a class of quantum error-correcting codes characterized by sparse parity check matrices. These codes are an extension of classical LDPC codes \cite{Gallager1962LDPC, Spielman1996LDPC} and they offer promising performance for fault-tolerant quantum computing due to their low overhead relative to topological codes \cite{gottesman2014faulttolerant, bravyi2024high, Xu2024}.

A CSS quantum LDPC code can be represented by a pair of sparse matrices, \(H_X\) and \(H_Z\), whose rows and columns each have only a small number of non-zero entries. This sparsity is crucial both for the practical implementation of stabilizer measurement and for decoding algorithms like belief propagation.

Here we describe an important class of quantum LDPC codes called hypergraph product (HGP) codes.
Constructed from two classical binary linear codes with parity-check matrices \(H_1\in\mathbb{F}_2^{m_1\times n_1}\) and \(H_2\in\mathbb{F}_2^{m_2\times n_2}\), the HGP code is defined by the check matrices
\begin{equation}
H_X = \begin{pmatrix}
H_1 \otimes I_{n_2} & I_{m_1} \otimes H_2^T
\end{pmatrix},
\end{equation}
\begin{equation}
H_Z = \begin{pmatrix}
I_{n_1} \otimes H_2 & H_1^T \otimes I_{m_2}
\end{pmatrix}.
\end{equation}

An extension of this construction called the \emph{lifted product construction} generalizes the HGP framework~\cite{Panteleev_2022_lifted}.
While HGP codes have binary entries in their parity-check matrices, lifted product codes replace these scalar entries with higher-dimensional objects such as \emph{circulant matrices}. This process is called \emph{lifting}, as it effectively "lifts" scalar entries to matrix entries, increasing the number of qubits and stabilizers

In the lifted product construction, the parity-check matrices become:
\begin{equation}
H_X = \begin{pmatrix}
\tilde{H}_1 \otimes I_{n_2} & I_{m_1} \otimes \tilde{H}_2^T
\end{pmatrix},
\end{equation}
\begin{equation}
H_Z = \begin{pmatrix}
I_{n_1} \otimes \tilde{H}_2 & \tilde{H}_1^T \otimes I_{m_2}
\end{pmatrix},
\end{equation}
where $\tilde{H}_1$ and $\tilde{H}_2$ are $m_1\times n_1$ and $m_2\times n_2$ 
matrices, respectively, whose elements are $L\times L$ binary circulant matrices. 

The circulant matrices form a commutative ring under matrix multiplication, which plays an essential role for maintaining the commutation relations required for the code to satisfy the commutativity constraint. In this work, we use quantum LDPC codes from the two aforementioned code families to evaluate the decoding performance of BPGD over the quantum erasure channel.

\section{Decoding Erasures with BPGD}
\label{subsec:bpgd}

\subsection{Belief Propagation Decoding}

BP is an iterative message-passing algorithm that operates on the Tanner graph of an LDPC code~\cite{kschischang2001factor}. For CSS codes, the correction of independent \( X \mkern-2mu \)- and \( Z \)-type errors can be treated separately.
Thus, we focus on the correction of \( X \mkern-2mu \)-type errors using the classical decoding with the parity-check matrix $H_Z$.
For correction of \( Z \)-type errors, one can  follow the same approach using classical decoding with the parity-check matrix $H_X$.
BP decoding of quantum LDPC codes was first introduced in~\cite{mackay2004sparse} and considered further in~\cite{poulin2008iterative}.

In this context, the Tanner graph has variable nodes \( V = \{v_1, \ldots, v_n\} \) representing elements of the \( X \)-error vector \( e = (e_1, \ldots, e_n) \) and check nodes \( C = \{c_1, \ldots, c_m\} \) representing the \( Z \)-stabilizers in the matrix \( H_Z \).
A variable node \( v_i \) is connected to a check node \( c_j \) if \( H_Z(i,j) = 1 \), meaning that the corresponding $j$-th qubit participates in the parity-check defined by the $i$-th $Z$-stabilizer.
The BP algorithm estimates the most likely values for the error bits by exchanging messages between the variable nodes and the check nodes (representing stabilizers).

Let $m^{(t)}_{v \to c}$ denote the BP message passed from variable node $v$ to check node $c$ during the $t$-th iteration and let $m^{(t)}_{c \to v}$ denote the BP message passed from check node $c$ to variable node $v$ during the $t$-th iteration.
The messages are typically represented by log-likelihood ratios (LLRs).
For a channel where $X\mkern-2mu$-errors occur with probability $p$, the variable nodes are initialized to the channel LLR defined by
\begin{equation}
 \lambda(p) = \ln \frac{1-p}{p}.
    \label{eq:v_prior}
\end{equation}

This LLR represents the initial belief that a qubit has an $X\mkern-2mu$-error.

At the start of the BP algorithm, these initial LLRs are used to set the messages sent from each variable node \( v \) to all its connected check nodes \( c \). Specifically, at iteration \( t = 0 \), the message from variable node \( v \) to check node \( c \) is set to
\begin{equation}
    m^{(0)}_{v \to c} = \mu_{v},
    \label{eq:v_prior}
\end{equation}
where $\mu_v = \lambda(p_v)$ and $p_v$ is the prior probability that qubit $v$ has an $X\mkern-2mu$-error.

Check nodes subsequently update their messages based on the incoming messages from all connected variable nodes and the syndrome bits $s = \{s_1, s_2, ..., s_m\}$, using the equation
\begin{equation}
    m^{(t)}_{c \to v} = (-1)^s 2\tanh^{-1} \left(\prod_{v' \in \cN(c) \setminus v} \tanh\left(\frac{m^{(t)}_{v' \to c}}{2}\right)\right),
    \label{check2bit}
\end{equation}
\noindent where \(\cN(c)\) represents the set of variable nodes connected to the check node \(c\), and \(m_{v' \to c}\) is the message from variable node \(v'\) to check node \(c\).

Then, the variable nodes update their beliefs using the messages received from check nodes:
\begin{equation}
    m^{(t+1)}_{v \to c} = \mu_v + \sum_{c' \in \cN(v) \setminus c} m^{(t)}_{c' \to v},
    \label{bit2check}
\end{equation}
\noindent where \(\mu_v\) is the initial LLR for variable node \(v\), and \(\cN(v)\) is the set of check nodes connected to \(v\).
Finally, the variable node output message is computed with
\begin{equation}
    m^{(t+1)}_{v} = \mu_v + \sum_{c' \in \cN(v)} m^{(t)}_{c' \to v}.
\end{equation}

\subsection{BP Convergence Issues for Quantum LDPC Codes}

For stabilizer codes, the commutativity constraint requires that all stabilizers commute.
When applied to CSS codes, this implies that the binary representations of all $X\mkern-2mu$-stabilizers must be codewords of the binary linear code with parity-check matrix $H_Z$.
In terms of erasure decoding, this means that the variable nodes in the support of an $X\mkern-2mu$-stabilizer must be a stopping set for the peeling decoder based on $H_Z$.
Moreover, the subgraph of the Tanner graph induced by a stopping set must contain ac cycle.
Thus, short cycles in the Tanner graph arise from the inherent degeneracy of CSS LDPC codes due to low-weight stabilizers and this affects BP convergence.

While BP convergence suffers, there is no reason that decoding performance must also suffer.
This is because the resulting uncertainty in the $X\mkern-2mu$-error pattern is exactly equal to an $X\mkern-2mu$-stabilizer.
This observation was highlighted in~\cite{connolly2024fast} and used to motivate the pruned peeling decoder.

The key idea of pruned peeling is that, if a $X\mkern-2mu$-stabilizer completely is covered by erasures, then any bit in its supprt can be fixed to 0 (i.e., no error) without risk because applying the stabilizer can always flip its value to 1.
We note that this operation ``uses up'' that stabilizer.
Pruned peeling makes use of this idea by searching for linear combinations of at most $M$ stabilizer generators which are completely covered by erasures and fixing one of their variables to 0.

\subsection{Belief Propagation with Guided Decimation}

Building on BP, the BPGD decoding algorithm operates by sequentially running BP and fixing values of variable nodes based on the resulting BP beliefs.
Message-passing algorithms that incorporate “decimation” were first introduced in the classical context for constraint satisfaction problems \cite{mezard2002satisfiability, Montanari2007SolvingCS}. 

For quantum LDPC codes, BPGD decoding seems to mitigate the non-convergence issue of BP due to the commutativity constraint of stabilizer codes.
 
In \cite{yao2023belief}, the BPGD decoding algorithm is shown to be effective for correcting bit-flip errors in quantum CSS codes. 

BPGD begins by initializing the LLRs for each variable node on the Tanner graph, similar to the standard BP algorithm. The process proceeds iteratively where, in each round, BP is run for a fixed number of iterations \( T \) using the sum-product algorithm to compute the estimated beliefs for the variable nodes.

If BP converges and the resulting error estimate matches the observed syndrome, the decoding process terminates, and the hard values of the variable nodes are returned as the estimated error. However, if convergence is not achieved, the algorithm identifies the variable node with the highest reliability, defined by the magnitude of the LLR, and decimates that bit. Decimation involves fixing the value of this variable node based on the sign of its current bias and marking the bit as decimated, effectively reducing the complexity of subsequent BP iterations.

The decimation step is controlled by a parameter \( {llr}_{\max} \), which is set to a large value (typically \( {llr}_{\max} = 25 \)) to ensure numerical stability in practical implementations. This fixed value serves as a strong bias, guiding the remaining variable nodes toward a solution in the subsequent rounds of BP. The process repeats, with each round consisting of BP followed by decimation, until either all variable nodes have been decimated or BP successfully converges.

While BPGD often enhances the probability of convergence compared to standard BP, there is a potential for \textit{non-convergence failure} if the final set of hard-decimated values does not match the syndrome. Nonetheless, BPGD seems to be a valuable approach, particularly in scenarios where traditional BP struggles to find a solution. The effectiveness of BPGD in such cases underscores its utility as a robust decoder for quantum error correction.

\subsection{BPGD Decoding for the Quantum Erasure Channel}

\begin{algorithm}[t]
\caption{BPGD over erasures}
\label{alg:BPGD_erasure}
\KwIn{
erasure locations $V_e$, block length $n$, $H_Z$ Tanner graph $G\!=\!(V,C,E)$,
syndrome $s$,
BP iterations per round $T$}
\KwOut{estimated $\widehat{x}$ or non-convergence}
\For{$i=1$ \emph{\KwTo} $n$}
{\eIf{$v_i \in V_e$}{
    $\mu_{v_i} = {llr}_{\min}$
}{
    $\mu_{v_i} = {llr}_{\max}$
}
$m^{(0)}_{v_i\rightarrow c_j} \leftarrow \mu_{v_i}$ for all 
$c_j\in \cN(v_i)$}

$V_u = V$ \\
\For{$r=1$ \emph{\KwTo} $n$}{
    run BP on Tanner graph $G$ for $T$ iterations\\ 
    $\widehat{x} \leftarrow$ hard values of the variable nodes\\
    
    \eIf{$\widehat{x}H_1^T = s$}{
        \Return{$\widehat{x}$}
    }{
        $v_i = \arg\max_{v\in V_u}\gamma(v_i)$ \\
        \eIf{$m_{v_i}^{(rT)} \geq 0$}{
            $\mu_{v_i} = {llr}_{\max}$
        }{
            $\mu_{v_i} = -{llr}_{\max}$
        }
        $V_u = V_u\backslash\{v_i\}$
    }
}
\Return{non-convergence}
\end{algorithm}
Here, we apply the BPGD algorithm to quantum erasure channels by choosing the initial LLRs correctly.
Specifically, when a bit is erased, the implied error rate is $1/2$ and the corresponding LLR value is $\lambda(1/2) = 0$. One can also use a very small value, ${llr}_{\min} \approx 0$, to avoid numerical issues.

Conversely, when a qubit is not erased, the decoder is very confident that the qubit is not in error. While the formal LLR value should be $\lambda(0)=\infty$, a large finite value is used for numerical reasons.
We define ${llr}_{\max} = 25$ to be this value.

These modifications align the decoder with the characteristics of erasure channels and result in reasonable decoding performance.
The pseudo-code for the BPGD algorithm, along with further technical details, can be found in Algorithm 1.

To compare BPGD with pruned peeling, consider the case where ${llr}_{\max} = \infty$ and BP iterates to a fixed point.
This is equivalent to the peeling decoder and always outputs a stopping set.
At this point, the pruned peeling decoder will make a guess only if there is a stabilizer that can absorb the possible error.
On the other hand, the BPGD decoder simply guesses bits in the stopping set randomly and continues decoding.
While this approach has no guarantees, some of the guesses will be correct and the others can hopefully be absorbed by the stabilizer.
Overall, performance is improved over pruned peeling.

\section{BPGD with Damping and Adjusted LLRs}

The performance of BPGD for quantum LDPC codes with erasures can be improved by adjusting the initial LLRS and incorporating damping into the belief propagation process.
In particular, we observed experimentally that the convergence rate of BPGD on quantum LDPC codes over erasures can be significantly improved by these modifications.
This, in  turn, reduces the failure rate especially at erasure rates where the rate of non-convergence dominates the rate of logical error. 

\subsection{BPGD with Adjusted LLRs}

It has been observed that, for BP decoding over both classical \cite{hagiwara2012fixed} and quantum LDPC codes \cite{miao2023quaternary}, adjusting the initialization of BP on the variable nodes can potentially improve performance.
In the content of BPGD decoding of quantum LDPC codes over erasures, we have also observed performance improvement upon adjusting the priors on the variable nodes.

In particular, we found that reducing the initial confidence in the known bits can improve the convergence of the decoder. To achieve this, we introduce a scaling factor \( c_{opt} \) and adjust the initial LLRs for non-erased bits to be
\begin{equation}
    {llr}_{\max}' = c_{opt} \;{llr}_{\max}.
    \label{v_llr_adjust}
\end{equation}

By choosing \( c_{opt} < 1 \), we reduce the initial LLR magnitude for non-erased bits and this decreases the number of decoder failures due to non-convergence.

To select the optimal value of \( c_{opt} \), we performed experiments over different ranges of values for different erasure rates, fine-tuning to narrower intervals when necessary. We observed that lower erasure rates required lower values of \( c_{opt} \) than higher erasure rates for to better convergence and lower failure rates. 
The optimized values of \( c_{opt} \) used in our simulations is provided in Table \ref{tab1}.

\begin{table}[t]
\centering
\caption{Optimized values of \( c_{opt} \) for simulations}
\label{tab1}
\centering
\setlength{\tabcolsep}{3pt}
\begin{tabular}{ccccccc}  
\hline\\[-3mm]
\multicolumn{7}{c}{\textbf{[[1600,64]] HGP code}} \\
\hline
\textbf{Erasure Rate} & $\leq0.08$ & 0.10 & 0.12 & 0.14-0.24 & 0.26 & $\geq0.28$ \\
\( c_{opt} \) &   0.1   &   0.2   &    0.1  &  0.2    &   0.3  &   0.5  \\
\hline\\[-3mm]
\multicolumn{7}{c}{\textbf{[[2025,81]] HGP code}} \\
\hline
\textbf{Erasure Rate} & $\leq0.12$ & 0.14-0.22 & $\geq0.24$ & & & \\
\( c_{opt} \) &  0.1  &  0.15  &  0.3  & & & \\
\hline\\[-3mm]
\multicolumn{7}{c}{\textbf{B1 code}} \\
\hline
\textbf{Erasure Rate} & $\leq0.34$ & 0.36-0.46 & 0.48 & 0.50 & & \\
\( c_{opt} \) &  0.3  &  0.5  &  0.8  &  0.9 & & \\
\hline
\end{tabular}
\vspace{-2mm}
\end{table}

\subsection{BPGD with Damping}

\begin{table}[t]
\centering
\caption{Optimized values of \( \gamma \) for simulations}
\label{tab2}
\setlength{\tabcolsep}{3pt} 
\begin{tabular}{cccccccc}  
\hline\\[-3mm]
\multicolumn{8}{c}{\textbf{[[1600,64]] HGP code}} \\
\hline
\textbf{Erasure Rate} & $\leq 0.06$ & 0.08-0.20 & 0.22-0.24 & 0.26 & 0.28 & 0.30  & 0.32 \\
\( \gamma \) &   0.87   &   0.88   &    0.90  &  0.93    &   0.94  &   0.95 & 0.96  \\
\hline\\[-3mm]
\multicolumn{8}{c}{\textbf{[[2025,81]] HGP code}} \\
\hline
\textbf{Erasure Rate} & $\leq 0.08$ & 0.10-0.16 & 0.18-0.22 & 0.24 & 0.26& 0.28 & 0.30 \\
\( \gamma \) &  0.86  &   0.87   &   0.88   &    0.90  &  0.92    &   0.95  &   0.96 \\
\hline
\textbf{Erasure Rate} & 0.32 &  &  & &  &  &  \\
\( \gamma \) &  0.97  &  &  & &  &  &   \\
\hline\\[-3mm]
\multicolumn{8}{c}{\textbf{B1 code}} \\
\hline
\textbf{Erasure Rate} & $\leq 0.36$ & 0.38-0.40 & 0.42-0.48 & 0.50 & & & \\
\( \gamma \) &  0.90  &  0.95  &  0.94  &  0.95 & & & \\
\hline
\end{tabular}
\vspace{-2mm}
\end{table}

During the BPGD decoding process, BP is applied to compute approximate marginals for the variable nodes that guide the decimation. 
This approximation is exact if the underlying Tanner graph is a tree \cite{kschischang2001factor}.
However, the Tanner graphs of quantum LDPC codes have many short cycles due to low-weight stabilizers and
BP may approximate the marginal poorly. If the main problem is that BP is not converging to a fixed point, then one technique that can be used to improve the BP performance is called \emph{damping} \cite{pretti2005message,nachmani2018deep,lian2019learned}.

At each step of the BP iteration in equation \eqref{bit2check}, the evaluation of \(m^{(t+1)}_{v \to c}\) is taken to be a weighted average between
the old estimate and the new estimate:
\begin{equation}
    \tilde{m}^{(t+1)}_{v \to c} = \mu_v + \sum_{c' \in N(v) \setminus c} m^{(t)}_{c' \to v}.
    \label{bit2check_damp1}
\end{equation}
The damped variable-to-check messages are computed with
\begin{equation}
    m^{(t+1)}_{v \to c} = (1 - \gamma) \cdot m^{(t)}_{v \to c} + \gamma \cdot \tilde{m}^{(t+1)}_{v \to c},
    \label{bit2check_damp2}
\end{equation}
\noindent where \( \gamma \in [0,1] \) is the damping factor.

By choosing an appropriate value for the damping factor \( \gamma \) we can affect the influence of new messages versus old messages.
A smaller \( \gamma \) provides more damping to help prevent oscillations and improves convergence of BP especially for lower erasure rates. The optimized values of \( \gamma \) used in our simulations are provided in Table \ref{tab2}.

\subsection{BPGD with Combined Adjustments}

For the [[2025, 81]] HGP code, neither adjusting the initial LLR nor the damping factor were enough by themselves for the BPGD performance to match the VH decoder. But, jointly optimizing LLR adjustment and damping factor provided further improvement. First, we performed a grid search over both parameters and found that the damping cofficient \( \gamma \) did not change much for a given erasure rate. Thus, we found suitable parameters by fixing \( \gamma \) for each erasure rate and searching over possible values of \( c_{opt} \).
The chosen values minimized the failure rates in our experiments.

Table \ref{tab3} summarizes the optimized values of \( c_{opt} \) used in our simulations for different erasure rates with the \( \gamma \) values being same as in Table \ref{tab2}.

\section{Numerical Results}

\begin{table}[t]
\centering
\caption{Optimized values of \( c_{opt} \) with \( \gamma \) values from Table \ref{tab2}}
\label{tab3}
\setlength{\tabcolsep}{4pt}
\begin{tabular}{cccccccc}
\hline\\[-3mm]
\multicolumn{8}{c}{\textbf{[[2025,81]] HGP code}} \\
\hline
Erasure Rate & 0.10 & 0.12 & 0.14 & 0.16 & 0.18 & 0.20 & 0.22 \\
\( c_{opt} \) &    0.065   & 0.077     &   0.077   &    0.089 &    0.101   &  0.11    &  0.115 \\
\hline
Erasure Rate & 0.24 & 0.26 & 0.28 & 0.30 & 0.32 &     &     \\
\( c_{opt} \)     &  0.135    & 0.14     &    0.19  &  0.23 &  0.25  &       &    \\
\hline \\[-1.5ex]
\end{tabular}
\vspace{-2mm}
\end{table}

\begin{figure}[]
    \centering
    \scalebox{1.0}{
\begin{tikzpicture}

\definecolor{brown}{RGB}{165,42,42}
\definecolor{darkgray176}{RGB}{176,176,176}
\definecolor{green}{RGB}{0,128,0}
\definecolor{lightgray204}{RGB}{204,204,204}
\definecolor{lightgreen}{RGB}{144,238,144}
\definecolor{orange}{RGB}{255,165,0}
\definecolor{purple}{RGB}{128,0,128}
\definecolor{yellow}{RGB}{255,255,0}

\begin{axis}[
legend cell align={left},
legend style={
  fill opacity=0.7,
  draw opacity=1,
  text opacity=1,
  at={(0.99,0.01)},
  anchor=south east,
  draw=lightgray204,
  font=\scriptsize
},
width=3.8in,
height=2.7in,
xlabel={Erasure Rate},
ylabel={Failure Rate},
grid=both,
ticklabel style = {font=\footnotesize},
label style = {font=\footnotesize},
x grid style={darkgray176},
xmin=0.03, xmax=0.32,
xtick={0.00,0.05,0.10,0.15,0.20,0.25,0.30,0.35,0.40,0.45,0.50},
xticklabels={0.00,0.05,0.10,0.15,0.20,0.25,0.30,0.35,0.40,0.45,0.50},
ymode=log,
y grid style={darkgray176},
ymin=1e-8, ymax=0.1,
ytick={1,1e-1,1e-2,1e-3,1e-4,1e-05,1e-06,1e-07,1e-08},
scale=0.88
]

\addplot+[
  thick, red, mark=*, mark options={scale=0.75},
  smooth, 
  error bars/.cd, 
    y dir=both, 
    y explicit
] table [x=x, y=y, y error plus=error1, y error minus=error2, col sep=comma] {
   x,            y,       error1,       error2
0.06, 0.0000008889, 0.0000005509, 0.0000005508
0.08, 0.0000048000, 0.0000011087, 0.0000011086
0.10, 0.0000165333, 0.0000029101, 0.0000029100
0.12, 0.0000465333, 0.0000048820, 0.0000048819
0.14, 0.0001202667, 0.0000078482, 0.0000078481
0.16, 0.0003066667, 0.0000396271, 0.0000396270
0.18, 0.0006026667, 0.0000555435, 0.0000555434
0.20, 0.0011640000, 0.0000771701, 0.0000771700
0.22, 0.0020866667, 0.0001032758, 0.0001032757
0.24, 0.0037533333, 0.0003094585, 0.0003094584
0.26, 0.0067866667, 0.0004154894, 0.0004154893
0.28, 0.0126000000, 0.0017850167, 0.0017850166
0.30, 0.0223333333, 0.0023647358, 0.0023647357
0.32, 0.0382000000, 0.0030674996, 0.0030674995
};
\addlegendentry{VH}

\addplot+[
  thick, blue, mark=*, mark options={scale=0.75},
  smooth, 
  error bars/.cd, 
    y dir=both, 
    y explicit
] table [x=x, y=y, y error plus=error1, y error minus=error2, col sep=comma] {
   x,            y,       error1,       error2
0.04, 0.0000002000, 0.0000002263, 0.0000001999
0.06, 0.0000018667, 0.0000006914, 0.0000006913
0.08, 0.0000098667, 0.0000015896, 0.0000015895
0.10, 0.0000352000, 0.0000030024, 0.0000030023
0.12, 0.0001056667, 0.0000052018, 0.0000052017
0.14, 0.0002595333, 0.0000081517, 0.0000081516
0.16, 0.0005767333, 0.0000121499, 0.0000121498
0.18, 0.0011691333, 0.0000172937, 0.0000172936
0.20, 0.0022014667, 0.0000237185, 0.0000237184
0.22, 0.0040038667, 0.0000319580, 0.0000319579
0.24, 0.0070222667, 0.0000422590, 0.0000422589
0.26, 0.0122264000, 0.0000556146, 0.0000556145
0.28, 0.0214063581, 0.0000814887, 0.0000814886
0.30, 0.0379079988, 0.0001500350, 0.0001500349
0.32, 0.0675776557, 0.0002757221, 0.0002757220
};
\addlegendentry{BPGD}

\addplot+[
  thick, green, mark=*, mark options={scale=0.75},
  smooth, 
  error bars/.cd, 
    y dir=both, 
    y explicit
] table [x=x, y=y, y error plus=error1, y error minus=error2, col sep=comma] {
   x,            y,       error1,       error2
0.04, 0.0000005333, 0.0000004268, 0.0000004267
0.06, 0.0000079111, 0.0000016436, 0.0000016435
0.08, 0.0000390667, 0.0000031630, 0.0000031629
0.10, 0.0001642667, 0.0000091720, 0.0000091719
0.12, 0.0005108000, 0.0000161711, 0.0000161710
0.14, 0.0014060000, 0.0000268171, 0.0000268170
0.16, 0.0034160000, 0.0001320509, 0.0001320508
0.18, 0.0075600000, 0.0001960372, 0.0001960371
0.20, 0.0152386667, 0.0002772454, 0.0002772453
0.22, 0.0292440000, 0.0003813282, 0.0003813281
0.24, 0.0537000000, 0.0011408065, 0.0011408064
0.26, 0.0917400000, 0.0014608139, 0.0014608138
0.28, 0.1565333333, 0.0058149746, 0.0058149745
0.30, 0.2472000000, 0.0069035819, 0.0069035818
0.32, 0.3628666667, 0.0076948320, 0.0076948319
};
\addlegendentry{Peeling}

\addplot+[
  thick, purple, mark=*, mark options={scale=0.75},
  smooth, 
  error bars/.cd, 
    y dir=both, 
    y explicit
] table [x=x, y=y, y error plus=error1, y error minus=error2, col sep=comma] {
   x,            y,       error1,       error2
0.04, 0.0000001778, 0.0000002464, 0.0000001777
0.06, 0.0000013333, 0.0000006748, 0.0000006747
0.08, 0.0000101333, 0.0000016110, 0.0000016109
0.10, 0.0000365333, 0.0000043258, 0.0000043257
0.12, 0.0001046667, 0.0000073216, 0.0000073215
0.14, 0.0002757333, 0.0000118826, 0.0000118825
0.16, 0.0007146667, 0.0000604814, 0.0000604813
0.18, 0.0015853333, 0.0000900412, 0.0000900411
0.20, 0.0032733333, 0.0001292732, 0.0001292731
0.22, 0.0066693333, 0.0001842102, 0.0001842101
0.24, 0.0137866667, 0.0005900999, 0.0005900998
0.26, 0.0257333333, 0.0008013041, 0.0008013040
0.28, 0.0502666667, 0.0034966433, 0.0034966432
0.30, 0.0898666667, 0.0045768035, 0.0045768034
0.32, 0.1514000000, 0.0057362080, 0.0057362079
};
\addlegendentry{Pruned peeling ($M\!=\!1$)}

\addplot+[
  thick, orange, mark=*, mark options={scale=0.75},
  smooth, 
  error bars/.cd, 
    y dir=both, 
    y explicit
] table [x=x, y=y, y error plus=error1, y error minus=error2, col sep=comma] {
   x,            y,       error1,       error2
0.04, 0.0000001778, 0.0000002464, 0.0000001777
0.06, 0.0000013333, 0.0000006748, 0.0000006747
0.08, 0.0000101333, 0.0000016110, 0.0000016109
0.10, 0.0000365333, 0.0000043258, 0.0000043257
0.12, 0.0001045333, 0.0000073170, 0.0000073169
0.14, 0.0002746667, 0.0000118596, 0.0000118595
0.16, 0.0007133333, 0.0000604250, 0.0000604249
0.18, 0.0015706667, 0.0000896244, 0.0000896243
0.20, 0.0032253333, 0.0001283250, 0.0001283249
0.22, 0.0065333333, 0.0001823348, 0.0001823347
0.24, 0.0134133333, 0.0005821654, 0.0005821653
0.26, 0.0249266667, 0.0007889713, 0.0007889712
0.28, 0.0483333333, 0.0034322289, 0.0034322288
0.30, 0.0859333333, 0.0044851834, 0.0044851833
0.32, 0.1456000000, 0.0056444515, 0.0056444514
};
\addlegendentry{Pruned peeling ($M\!=\!2$)}

\addplot+[
  thick, lightgreen, mark=*, mark options={scale=0.75},
  smooth, 
  error bars/.cd, 
    y dir=both, 
    y explicit
] table [x=x, y=y, y error plus=error1, y error minus=error2, col sep=comma] {
   x,            y,       error1,       error2
0.04, 0.0000000533, 0.0000000370, 0.0000000369
0.06, 0.0000007667, 0.0000001401, 0.0000001400
0.08, 0.0000043667, 0.0000003344, 0.0000003343
0.10, 0.0000180667, 0.0000021510, 0.0000021509
0.12, 0.0000507333, 0.0000036045, 0.0000036044
0.14, 0.0001252667, 0.0000056637, 0.0000056636
0.16, 0.0002908000, 0.0000086287, 0.0000086286
0.18, 0.0006116216, 0.0000125960, 0.0000125959
0.20, 0.0012000000, 0.0001752025, 0.0001752024
0.22, 0.0023200000, 0.0002434726, 0.0002434725
0.24, 0.0043333333, 0.0003324134, 0.0003324133
0.26, 0.0077866667, 0.0004448246, 0.0004448245
0.28, 0.0137333333, 0.0005889733, 0.0005889732
0.30, 0.0242733333, 0.0007788238, 0.0007788237
0.32, 0.0437466667, 0.0010350692, 0.0010350691
};
\addlegendentry{Damped BPGD}

\addplot+[
  thick, cyan, mark=*, mark options={scale=0.75},
  smooth, 
  error bars/.cd, 
    y dir=both, 
    y explicit
] table [x=x, y=y, y error plus=error1, y error minus=error2, col sep=comma] {
   x,            y,       error1,       error2
0.04, 0.0000000667, 0.0000000924, 0.0000000666
0.06, 0.0000008333, 0.0000003267, 0.0000003266
0.08, 0.0000047333, 0.0000007785, 0.0000007784
0.10, 0.0000171333, 0.0000014812, 0.0000014811
0.12, 0.0000476667, 0.0000078125, 0.0000078124
0.14, 0.0001236667, 0.0000125833, 0.0000125832
0.16, 0.0003033333, 0.0000197056, 0.0000197055
0.18, 0.0006433333, 0.0000907348, 0.0000907347
0.20, 0.0011600000, 0.0001218070, 0.0001218069
0.22, 0.0024466667, 0.0001767872, 0.0001767871
0.24, 0.0049600000, 0.0002513950, 0.0002513949
0.26, 0.0096019110, 0.0004937286, 0.0004937285
0.28, 0.0183453447, 0.0007103826, 0.0007103825
0.30, 0.0339068759, 0.0012940577, 0.0012940576
0.32, 0.0638225159, 0.0025567494, 0.0025567493
};
\addlegendentry{Adjusted LLR BPGD}

%

\addplot+[
  thick, black, mark=*, mark options={scale=0.75},
  smooth, 
  error bars/.cd, 
    y dir=both, 
    y explicit
] table [x=x, y=y, y error plus=error1, y error minus=error2, col sep=comma] {
   x,            y,       error1,       error2
0.04, 0.0000000533, 0.0000000370, 0.0000000369
0.06, 0.0000007667, 0.0000001401, 0.0000001400
0.08, 0.0000043067, 0.0000003321, 0.0000003320
0.10, 0.0000188000, 0.0000021942, 0.0000021941
0.12, 0.0000511333, 0.0000036187, 0.0000036186
0.14, 0.0001306667, 0.0000182921, 0.0000182920
0.16, 0.0002966667, 0.0000275601, 0.0000275600
0.18, 0.0006113333, 0.0000395564, 0.0000395563
0.20, 0.0011193333, 0.0000535115, 0.0000535114
0.22, 0.0020113333, 0.0000716994, 0.0000716993
0.24, 0.0035293333, 0.0000949050, 0.0000949049
0.26, 0.0058606667, 0.0001221540, 0.0001221539
0.28, 0.0099533333, 0.0005023687, 0.0005023686
0.30, 0.0156866667, 0.0006288433, 0.0006288432
0.32, 0.0236733333, 0.0007693744, 0.0007693743
};
\addlegendentry{ML}

\end{axis}

\end{tikzpicture}
    }
    \vspace{-6.5mm}
t    \caption{Comparison the quantum erasure channel of BPGD decoding with peeling, pruned peeling, VH, and ML decoding. The plot presents the decoder failure rates for the HGP QLDPC code [[1600, 64]] from~\cite{connolly2024fast}, with convergence to a degenerate codeword considered a success. The number of simulations per data point varies and was chosen to ensure short error bars.}
    \label{fig:BPGD_comparison_1600}
    \vspace{-2mm}
\end{figure}
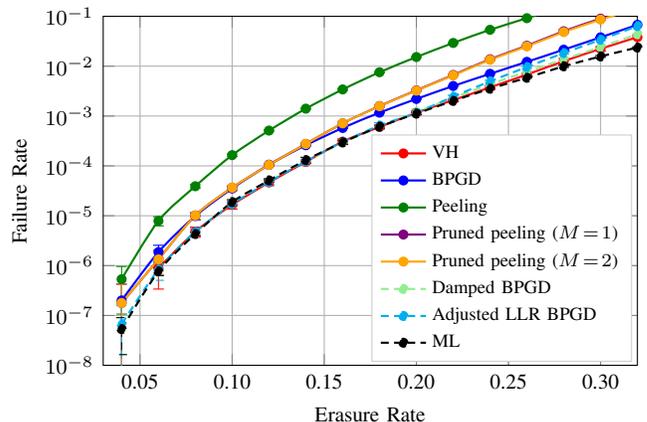

\begin{figure}[t]
    \centering
    \scalebox{1.0}{
    \input{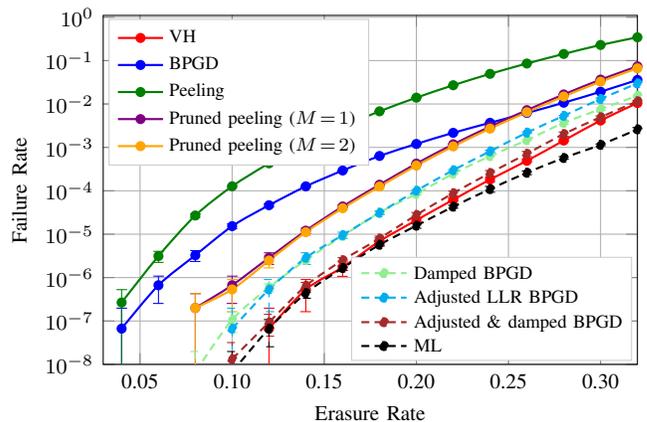}
    }
  \vspace{-6.5mm}
  \caption{Comparison on the quantum erasure channel of BPGD decoding with peeling, pruned peeling, VH, and ML decoding. The plot presents the decoder failure rates for the HGP QLDPC code [[2025, 81]]  from~\cite{connolly2024fast}.}
    \label{fig:BPGD_comparison_2025}
    \vspace{-2mm}
\end{figure}

In Figures \ref{fig:BPGD_comparison_1600} and \ref{fig:BPGD_comparison_2025}, we show simulation results for the BPGD decoder on the quantum erasure channel for two HGP QLDPC codes. These figures present the performance of peeling, pruned peeling, and VH decoders for comparison. The data points for the latter decoders were obtained by executing the code available in the publicly accessible GitHub repository referenced by~\cite{connolly2024fast}. BPGD decoding performs better than the peeling decoder for both codes but performs worse than pruned peeling for the [[2025, 81]] code.
BPGD with adjustments outperforms pruned peeling decisively, achieving proximity to VH decoder which itself is close to the ML decoding performance for most data points.

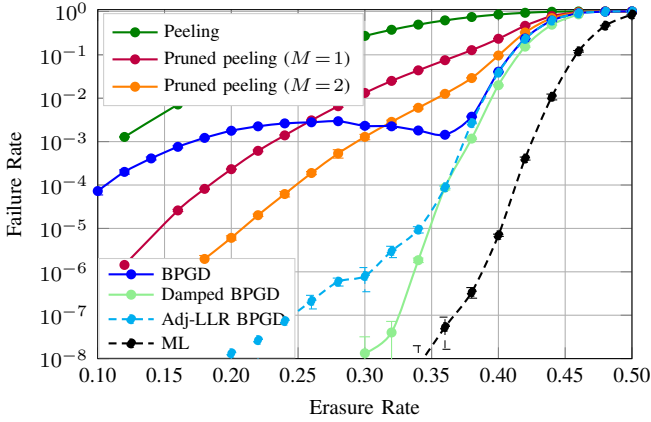
\begin{figure}[t]
    \centering
    \scalebox{1.0}{
\begin{tikzpicture}

\definecolor{darkgray176}{RGB}{176,176,176}
\definecolor{lightgray204}{RGB}{204,204,204}
\definecolor{lightgreen}{RGB}{144,238,144}
\definecolor{yellow}{RGB}{255,255,0}
\definecolor{green}{RGB}{0,128,0}

\begin{axis}[
legend cell align={left},
legend style={
  fill opacity=0.7,
  draw opacity=1,
  text opacity=1,
  at={(0.01,0.99)},
  anchor=north west,
  draw=lightgray204,
  font=\scriptsize
},
width=3.8in,
height=2.7in,
xlabel={Erasure Rate},
ylabel={Failure Rate},
grid=both,
ticklabel style = {font=\footnotesize},
label style = {font=\footnotesize},
x grid style={darkgray176},
xmin=0.1, xmax=0.5,
xtick={0.00,0.05,0.10,0.15,0.20,0.25,0.30,0.35,0.40,0.45,0.50},
xticklabels={0.00,0.05,0.10,0.15,0.20,0.25,0.30,0.35,0.40,0.45,0.50},
ymode=log,
y grid style={darkgray176},
ymin=1e-8, ymax=1.10,
ytick={1,1e-1,1e-2,1e-3,1e-4,1e-05,1e-06,1e-07,1e-08},
scale=0.88
]

\addplot+[
  thick, green, mark=*, mark options={scale=0.75},
  smooth, 
  error bars/.cd, 
    y dir=both, 
    y explicit
] table [x=x, y=y, y error plus=error1, y error minus=error2, col sep=comma] {
   x,            y,       error1,       error2
0.12, 0.0012803133, 0.0000057226, 0.0000057226
0.16, 0.0071674021, 0.0000237413, 0.0000237413
0.18, 0.0145025155, 0.0000336461, 0.0000336461
0.2,  0.0271782474, 0.0000457628, 0.0000457628
0.22, 0.0477916667, 0.0002414001, 0.0002414000
0.24, 0.0792880000, 0.0003057462, 0.0003057461
0.26, 0.1257083333, 0.0003751506, 0.0003751505
0.28, 0.1894600000, 0.0019831540, 0.0019831539
0.30, 0.2727400000, 0.0022538744, 0.0022538743
0.32, 0.3776133333, 0.0024533767, 0.0024533766
0.34, 0.4964466667, 0.0025302852, 0.0025302851
0.36, 0.6218933333, 0.0024540057, 0.0024540056
0.38, 0.7431066667, 0.0022111206, 0.0022111205
0.40, 0.8492066667, 0.0018109573, 0.0018109572
0.42, 0.9298000000, 0.0040885919, 0.0040885918
0.44, 0.9826000000, 0.0020925384, 0.0020925383
0.46, 0.9979333333, 0.0007267690, 0.0007267689
0.48, 1.0000000000, 0.0000000000, 0.0000000000
0.50, 1.0000000000, 0.0000000000, 0.0000000000
};
\addlegendentry{Peeling}

\addplot+[
  thick, purple, mark=*, mark options={scale=0.75},
  smooth, 
  error bars/.cd, 
    y dir=both, 
    y explicit
] table [x=x, y=y, y error plus=error1, y error minus=error2, col sep=comma] {
   x,            y,       error1,       error2
0.12, 0.0000014400, 0.0000001920, 0.0000001920 
0.16, 0.0000259793, 0.0000014345, 0.0000014345
0.18, 0.0000815463, 0.0000025414, 0.0000025414
0.2,  0.0002317113, 0.0000042836, 0.0000042836  
0.22, 0.0006136667, 0.0000280239, 0.0000280238
0.24, 0.0013926667, 0.0000422004, 0.0000422003
0.26, 0.0030890000, 0.0000627961, 0.0000627960
0.28, 0.0065466667, 0.0004081260, 0.0004081259
0.30, 0.0132266667, 0.0005781551, 0.0005781550
0.32, 0.0250733333, 0.0007912295, 0.0007912294
0.34, 0.0438866667, 0.0010366483, 0.0010366482
0.36, 0.0752066667, 0.0013346302, 0.0013346301
0.38, 0.1281133333, 0.0016913661, 0.0016913660
0.40, 0.2343866667, 0.0021437870, 0.0021437869
0.42, 0.4660666667, 0.0079832178, 0.0079832177
0.44, 0.7801333333, 0.0066278788, 0.0066278787
0.46, 0.9572000000, 0.0032391695, 0.0032391694
0.48, 0.9972666667, 0.0008355307, 0.0008355306
0.50, 1.0000000000, 0.0000000000, 0.0000000000
};
\addlegendentry{Pruned peeling ($M\!=\!1$)}

\addplot+[
  thick, orange, mark=*, mark options={scale=0.75},
  smooth, 
  error bars/.cd, 
    y dir=both, 
    y explicit
] table [x=x, y=y, y error plus=error1, y error minus=error2, col sep=comma] {
   x,            y,       error1,       error2
0.18,  0.0000019794, 0.0000003959, 0.0000003959
0.2,   0.0000060618, 0.0000006929, 0.0000006929
0.22, 0.0000201649, 0.0000012638, 0.0000012638
0.24, 0.0000620000, 0.0000089100, 0.0000089099
0.26, 0.0001890000, 0.0000155555, 0.0000155554
0.28, 0.0005333333, 0.0001168406, 0.0001168405
0.30, 0.0012866667, 0.0001814111, 0.0001814110
0.32, 0.0028533333, 0.0002699394, 0.0002699393
0.34, 0.0060200000, 0.0003914691, 0.0003914690
0.36, 0.0126133333, 0.0005647666, 0.0005647665
0.38, 0.0291400000, 0.0008512037, 0.0008512036
0.40, 0.0964000000, 0.0014936095, 0.0014936094
0.42, 0.3267333333, 0.0075058645, 0.0075058644
0.44, 0.7005333333, 0.0073299175, 0.0073299174
0.46, 0.9383333333, 0.0038495880, 0.0038495879
0.48, 0.9959333333, 0.0010184621, 0.0010184620
0.50, 1.0000000000, 0.0000000000, 0.0000000000
};
\addlegendentry{Pruned peeling ($M\!=\!2$)}

\addplot+[
  thick, blue, mark=*, mark options={scale=0.75},
  smooth, 
  error bars/.cd, 
    y dir=both, 
    y explicit
] table [x=x, y=y,y error=error, col sep=comma] {
   x,            y,        error
0.04, 0.0000002667, 0.0000002613
0.06, 0.0000049333, 0.0000011240
0.08, 0.0000232667, 0.0000024410
0.10, 0.0000726667, 0.0000136415
0.12, 0.0002006667, 0.0000226675
0.14, 0.0004100000, 0.0000323976
0.16, 0.0007586667, 0.0000440627
0.18, 0.0012193333, 0.0000558479
0.20, 0.0017786667, 0.0000674328
0.22, 0.0022513333, 0.0000758474
0.24, 0.0026300000, 0.0000819627
0.26, 0.0027893333, 0.0000844023
0.28, 0.0029540000, 0.0000868507
0.30, 0.0023133333, 0.0002431233
0.32, 0.0022533333, 0.0002399569
0.34, 0.0018033333, 0.0000678979
0.36, 0.0014373333, 0.0000606285
0.38, 0.0037513333, 0.0000978334
0.40, 0.0408505449, 0.0006400821
0.42, 0.2379668543, 0.0033238504
0.44, 0.6159625570, 0.0061081278
0.46, 0.9041335262, 0.0044792368
0.48, 0.9913428496, 0.0014760713
0.50, 0.9996669109, 0.0002919168
};
\label{bpgd}

\addplot+[
  thick, lightgreen, mark=*, mark options={scale=0.75},
  smooth, 
  error bars/.cd, 
    y dir=both, 
    y explicit
] table [x=x, y=y, y error plus=error1, y error minus=error2, col sep=comma] {
   x,            y,       error1,       error2
0.30, 0.0000000133, 0.0000000185, 0.0000000132
0.32, 0.0000000400, 0.0000000320, 0.0000000319
0.34, 0.0000018533, 0.0000002179, 0.0000002178
0.36, 0.0000873333, 0.0000105747, 0.0000105746
0.38, 0.0011733333, 0.0000387393, 0.0000387392
0.40, 0.0199266667, 0.0007072241, 0.0007072240
0.42, 0.1544000000, 0.0057825125, 0.0057825124
0.44, 0.4942666667, 0.0113153213, 0.0113153212
0.46, 0.8537333333, 0.0079975919, 0.0079975918
};
\label{bpgd_d}

\addplot+[
  thick, cyan, mark=*, mark options={scale=0.75},
  smooth, 
  error bars/.cd, 
    y dir=both, 
    y explicit
] table [x=x, y=y, y error plus=error1, y error minus=error2, col sep=comma] {
   x,            y,       error1,       error2
0.20, 0.0000000133, 0.0000000185, 0.0000000132
0.22, 0.0000000267, 0.0000000261, 0.0000000260
0.24, 0.0000000733, 0.0000000433, 0.0000000432
0.26, 0.0000002133, 0.0000000739, 0.0000000738
0.28, 0.0000005933, 0.0000001233, 0.0000001232
0.30, 0.0000008000, 0.0000004526, 0.0000004525
0.32, 0.0000030000, 0.0000008765, 0.0000008764
0.34, 0.0000094000, 0.0000015516, 0.0000015515
0.36, 0.0000887333, 0.0000047669, 0.0000047668
0.38, 0.0026666667, 0.0000825306, 0.0000825305
0.40, 0.0389586667, 0.0003096589, 0.0003096588
0.42, 0.2328882985, 0.0010320290, 0.0010320289
0.44, 0.6162354197, 0.0019316578, 0.0019316577
0.46, 0.9056625415, 0.0014076207, 0.0014076206
0.48, 0.9897674419, 0.0005066418, 0.0005066417
0.50, 0.9995403279, 0.0001084376, 0.0001084375
};
\label{bpgd_a}

\addplot+[
  thick, black, mark=*, mark options={scale=0.75},
  smooth, 
  error bars/.cd, 
    y dir=both, 
    y explicit
] table [x=x, y=y, y error plus=error1, y error minus=error2, col sep=comma] {
   x,            y,       error1,       error2
0.34, 0.0000000067, 0.0000000131, 0.0000000066
0.36, 0.0000000533, 0.0000000370, 0.0000000369
0.38, 0.0000003400, 0.0000000933, 0.0000000932
0.40, 0.0000069600, 0.0000004222, 0.0000004221
0.42, 0.0004080000, 0.0000323186, 0.0000323185
0.44, 0.0108000000, 0.0016541099, 0.0016541098
0.46, 0.1204666667, 0.0052091881, 0.0052091880
0.48, 0.4699333333, 0.0079871863, 0.0079871862
0.50, 0.8470000000, 0.0057609998, 0.0057609997
};
\label{yml}

\node [
    fill=white,
    fill opacity=0.7,
    draw opacity=1,
    text opacity=1,
    anchor=south west,
    draw=lightgray204,
    font=\scriptsize,
] at (rel axis cs: 0.001,0.001) {
\shortstack[l]{
\ref{bpgd} BPGD\\
\ref{bpgd_d} Damped BPGD\\
\ref{bpgd_a} Adj-LLR BPGD\\
\ref{yml} ML
}
};

\end{axis}

\end{tikzpicture}
    }
    \vspace{-6mm} 
    \caption{Comparison of BPGD decoding over the quantum erasure channel with peeling, pruned peeling, and ML decoding. The curves show the decoder failure rates for the [[882, 24, $18 \leq d \leq 24$]] B1 lifted-product QLDPC code from~\cite{Panteleev2021degeneratequantum}.}
    \label{fig:BPGD_comparison_B1}
\end{figure}

In Figure \ref{fig:BPGD_comparison_B1}, we show the simulation results for the [[882, 24, $18 \leq d \leq 24$]] B1 lifted-product code\footnote{This was called a generalized hypergraph product code in the original paper but the literature now uses the term lifted-product code.} from~\cite{Panteleev2021degeneratequantum}.
Adjusted versions of BPGD decoding are compared with peeling, pruned peeling, and ML decoding, all on the quantum erasure channel.
While the original BPGD decoder does not uniformly outperform pruned peeling due to convergence issues, BPGD with the proposed modifications is much closer to ML decoding than all other decoders.
We also want to highlight the sharp threshold-like behavior of both BPGD and ML decoding for the B1 code with erasures.
While this type of performance is typical for classical LDPC codes, the performance of quantum LDPC codes usually has the slower decay rate shown, for example, by pruned peeling with $M=2$.

\section{Conclusion}
\label{sec:conclusion}

In this paper, we present and analyze the performance of the belief propagation with guided decimation (BPGD) decoder on the quantum erasure channel. Specifically, we compare the performance of BPGD decoding with other established erasure decoders via simulation for a lifted product QLDPC code and two hypergraph product (HGP) QLDPC codes. The impressive performance of BPGD decoding for the [[882, 24, $18 \leq d \leq 24$]] B1 QLDPC code demonstrates the potential of BPGD as a generic decoder for QLDPC codes.

For the tested HGP codes, our results show that the BPGD decoder outperforms both peeling and pruned peeling decoders when correcting erasures. Although the BPGD decoder without adjustment is slightly inferior to the Vertical-Horizontal (VH) decoder for these codes, its computational complexity is substantially lower.

Moreover, we observe that small adjustments to the BP algorithm can enhance BPGD decoding performance by encouraging the decoder to converge.
In particular, by adjusting the initial LLRs and applying damping techniques, we find that BPGD decoding can approach the performance of the VH decoder.

Finally, we note that quantum codes with erasure conversion are likely to have both erasures and errors.
Due to its good performance for channels with erasures and channels with erros, the BPGD decoder is an ideal candidate for channels with both errors and erasures.

\section*{Acknowledgements}

This material is based on work supported by the NSF under Grants 2106213 and 2120757. 
Any opinions, findings, and conclusions or recommendations expressed in this material are those of the authors and do not necessarily reflect the views of the NSF.

The decoder performance curves from~\cite{connolly2024fast} have been reproduced in this work based on a larger simulation study.
We also acknowledge the use of generative AI tools for assistance in coding-related tasks during the development of our decoding algorithms. 

\nocite{roffe_decoding_2020}
\nocite{gr_ldpc_repo}

\bibliographystyle{IEEEtran}
\IEEEtriggeratref{37}
\bibliography{refs}

\end{document}